\documentclass[12pt]{article}
\usepackage{graphicx}
\usepackage{amsmath}
\usepackage{amssymb}
\usepackage{caption2}
\begin{document}
\bibliographystyle{prsty}
\begin{center}
{\large {\bf \sc{  Analysis of  $\Omega_b^-(bss)$  and  $\Omega_c^0(css)$ with QCD sum rules }}} \\[2mm]
Zhi-Gang Wang \footnote{E-mail,wangzgyiti@yahoo.com.cn.  }     \\
 Department of Physics, North China Electric Power University,
Baoding 071003, P. R. China

\end{center}

\begin{abstract}
In this article, we calculate the masses and the pole residues
  of the $\frac{1}{2}^+$ heavy  baryons $\Omega_c^0(css)$ and $\Omega_b^-(bss)$
   with the QCD sum rules.  The  numerical values
   $M_{\Omega_c^0}=(2.72\pm0.18)\,\rm{GeV}$ (or $M_{\Omega_c^0}=(2.71\pm0.18)\,\rm{GeV}$)
and $M_{\Omega_b^-}=(6.13\pm0.12)\,\rm{GeV}$ (or
$M_{\Omega_b^-}=(6.18\pm0.13)\,\rm{GeV}$)
   are in good agreement
    with  the experimental data.
\end{abstract}



\section{Introduction}

The charmed and bottomed baryons which  contain a heavy quark and
two light quarks are particularly interesting for studying dynamics
of the light quarks in the presence  of a heavy quark. They behave
as the QCD analogue of the familiar hydrogen bounded by the
electromagnetic interaction, and serve as an excellent ground for
testing predictions of the constituent  quark models and heavy quark
symmetry \cite{ReviewH1,ReviewH2}. The $u$, $d$ and $s$ quarks form
an $SU(3)$ flavor triplet, ${\bf 3}\times {\bf 3}={\bf \bar 3}+{\bf
6}$, two light quarks can form diquarks of a symmetric sextet  and
an antisymmetric antitriplet. For the $S$-wave baryons, the sextet
contains both spin-$\frac{1}{2}$ and spin-$\frac{3}{2}$ states,
while the antitriplet contains only spin-$\frac{1}{2}$ states.

There have been many theoretical approaches dealing  with  the heavy
baryons with one heavy quark,  for example,  the relativized
potential quark  model \cite{Capstick86}, the Feynman-Hellmann
theorem and semi-empirical mass formulas \cite{Roncaglia95}, the
Faddeev equation \cite{Sil96,Valcarce08}, the combined expansion in
$1/m_Q$ and $1/N_c$ \cite{Jenkins96}, the lattice simulations
\cite{Bowler96,Mathur02}, the relativistic quark-diquark
approximation \cite{Ebert05,Ebert08}, the hyperfine interactions
\cite{Karliner07}, the QCD sum rules in the heavy quark effective
field theory \cite{Liu07}, the variational approach
\cite{Roberts07}, and the full QCD sum rules
 \cite{Bagan921,Bagan922}, etc.

The ${1\over 2}^+$ antitriplet states ($\Lambda_c^+$,
$\Xi_c^+,\Xi_c^0)$,  and the ${1\over 2}^+$ and ${3\over 2}^+$
sextet states  ($\Omega_c,\Sigma_c,\Xi'_c$) and
($\Omega_c^*,\Sigma_c^*,\Xi'^*_c$) have been established; while the
corresponding bottomed  baryons are far from complete, only the
$\Lambda_b$, $\Sigma_b$, $\Sigma_b^*$, $\Xi_b$ have been  observed
\cite{PDG}.

Recently the D0 collaboration reported the observation of the doubly
strange baryon $\Omega_{b}^{-}$ in the decay channel $\Omega_b^- \to
J/\psi\thinspace\Omega^-$ (with $J/\psi\to\mu^+\mu^-$ and
$\Omega^-\to\Lambda K^-\to p\pi^- K^-$) in $p\bar{p}$ collisions at
$\sqrt{s}=1.96$ TeV \cite{OmegaB}. The experimental value
$M_{\Omega_b^-}=6.165\pm 0.010\thinspace \pm 0.013\thinspace \,
\rm{GeV}$ is about $0.1 \, \rm{GeV}$ larger than the existing
theoretical calculations
\cite{Roncaglia95,Valcarce08,Jenkins96,Bowler96,Mathur02,Ebert05,Ebert08,Karliner07,Liu07,Roberts07},
while the theoretical prediction $M_{\Omega_c^0}\approx 2.7\,
\rm{GeV}$
\cite{Roncaglia95,Valcarce08,Jenkins96,Bowler96,Mathur02,Ebert05,Ebert08,Karliner07,Liu07,Roberts07}
is consistent with the experimental data
$M_{\Omega_c^0}=(2.6975\pm0.0026) \,\rm{GeV}$ \cite{PDG} .

In previous work, we  have calculated  the masses and  the pole
residues of the $\frac{3}{2}^+$ heavy baryons  $\Omega_c^*$ and
$\Omega_b^*$
 with the QCD sum rules \cite{Wang0704}. In this article, we extend our previous work to
study the corresponding $\frac{1}{2}^+$ heavy baryons
 $\Omega_c^0$ and $\Omega_b^-$.

 In Ref.\cite{Nielsen07}, the masses of the heavy
baryons $\Omega_c^0$ and $\Omega_b^-$ are studied in the full QCD
sum rules with the same interpolating current as the present work.
The authors choose the tensor structure $\!\not\!{p}$ and obtain
different spectral density \footnote{When I finish the work and
submit it to the network http://arXiv.org, I learn from Prof.
Nielsen   that they have studied the masses of the heavy baryons
$\Omega_c^0$ and $\Omega_b^-$ in the full QCD sum rules with the
same interpolating current $J^a(x)$ \cite{Nielsen07}.}.

 The masses of the $\Lambda_{Q}$, $\Sigma_{Q}$ and $\Xi_{Q}$
 have been calculated with the full QCD sum rules
 \cite{Bagan921,Bagan922,Nielsen07}.    The masses
of the $\Sigma^*_Q$, $\Sigma_Q$ and $\Lambda_Q$ have been calculated
with the QCD sum rules in the leading order of the heavy quark
effective theory  \cite{Shuryak82,Grozin92,Bagan93}, and later the
$1/m_Q$ corrections were  studied \cite{Dai961,Dai962,Huang02}.
Furthermore, the masses of the orbitally excited heavy baryons with
the leading order approximation \cite{Zhu00,HuangCS}
 and the $1/m_Q$ corrections \cite{HuangMQ} in the heavy
quark effective theory have also been analyzed. Recently the
$\frac{1}{2}^+$ and $\frac{3}{2}^+$  bottomed  baryons were studied
with the QCD sum rules in the heavy quark effective theory including
the $1/m_Q$ corrections \cite{Liu07}.

  In the QCD sum rules, the operator product
expansion is used to expand the time-ordered currents into a series
of quark and gluon condensates which parameterize the long distance
properties of the QCD vacuum. Based on the quark-hadron duality, we
can obtain copious information about the hadronic parameters at the
phenomenological side \cite{SVZ79,PRT85}.

The article is arranged as follows:  we derive the QCD sum rules for
the masses and the pole residues of  the $\Omega_c^0$ and
 $\Omega_b^-$  in section 2; in section 3 numerical results are given and
 discussed,  and  section 4 is reserved for conclusion.

\section{QCD sum rules for  the $\Omega_c^0$ and $\Omega_b^-$}
In the following, we write down  the two-point correlation functions
$\Pi^a(p)$ in the QCD sum rules approach,
\begin{eqnarray}
\Pi^a(p)&=&i\int d^4x e^{ip \cdot x} \langle
0|T\left\{J^{a}(x)\bar{J}^{a}(0)\right\}|0\rangle \, ,  \\
J^{a}(x)&=&\epsilon^{ijk}s^T_i(x)C\gamma_\mu s_j(x)\gamma_5 \gamma^\mu Q^a_k(x) \, , \\
\lambda_{a}N(p,s)  &=& \langle 0|J^{a}(0)|\Omega_a(p,s)\rangle\, ,
\end{eqnarray}
where the upper index $a$ represents the $c$ and $b$ quarks
respectively,  $i$, $j$ and $k$ are color indexes, $C$ is the charge
conjunction matrix, the $N(p,s)$ and $\lambda_{a}$ stand for the
Dirac spin vector and the pole residue of the heavy baryon
$\Omega_a$, respectively.

The correlation functions  $\Pi^a(p)$ can be decomposed as follows:
\begin{eqnarray}
\Pi^a(p)&=&\!\not\!{p}\Pi^a_1(p) +\Pi_2^a(p) \, ,
\end{eqnarray}
due to  Lorentz covariance.  The first structure $\!\not\!{p}$ has
an odd number of $\gamma$-matrices  and conserves chirality, the
second  structure $1$ has an even number of $\gamma$-matrices and
violate chirality. In the original QCD sum rules analysis of the
nucleon masses and magnetic moments
\cite{Ioffe81,Ioffe84,Ioffe82,Ioffe83}, the interval of dimensions
(of the condensates) for the odd structure is larger than the
interval of dimensions for the even structure, one may expect a
better accuracy of  the results obtained from the sum rules with
the odd structure.

In this article, we choose the two tensor structures  to study the
masses and the pole residues of the  heavy baryons $\Omega_c^0$ and
$\Omega_b^-$, as the masses of the heavy quarks break the chiral
symmetry explicitly.

The components $\Pi^a_1(p)$ with the odd structure have smaller
dimension of mass than  the components $\Pi^a_2(p)$ with the even
structure, naively we  expect the components $\Pi^a_1(p)$ have
better convergent behavior in the operator product expansion, and
the numerical results confirm this conjecture, see Tables 1-2.

 We  insert  a
complete set  of intermediate baryon states with the same quantum
numbers as the current operators $J^{a}(x)$ into the correlation
functions $\Pi^{a}(p)$  to obtain the hadronic representation
\cite{SVZ79,PRT85}. After isolating the pole terms  of the lowest
states $\Omega_a$, we obtain the following result:
\begin{eqnarray}
\Pi^a(p)&=&\lambda_{a}^2\frac{M_{\Omega_a}+\!\not\!{p}}{M_{\Omega_a}^2-p^2}
+\cdots \, \, .
\end{eqnarray}

In the following, we briefly outline  the operator product expansion
for the correlation functions $\Pi^a(p)$  in perturbative QCD. The
calculations are performed at   large space-like momentum region
$p^2\ll 0$, which corresponds to small distance $x\approx 0$
required by   validity of  the operator product expansion. We write
down the "full" propagators $S_{ij}(x)$ and $S_Q^{ij}(x)$ of a
massive quark in the presence of the vacuum condensates firstly
\cite{PRT85},
\begin{eqnarray}
S_{ij}(x)&=& \frac{i\delta_{ij}\!\not\!{x}}{ 2\pi^2x^4}
-\frac{\delta_{ij}m_s}{4\pi^2x^2}-\frac{\delta_{ij}}{12}\langle
\bar{s}s\rangle +\frac{i\delta_{ij}}{48}m_s
\langle\bar{s}s\rangle\!\not\!{x}-\frac{\delta_{ij}x^2}{192}\langle \bar{s}g_s\sigma Gs\rangle\nonumber\\
&& +\frac{i\delta_{ij}x^2}{1152}m_s\langle \bar{s}g_s\sigma
 Gs\rangle \!\not\!{x}-\frac{i}{32\pi^2x^2} G^{ij}_{\mu\nu} (\!\not\!{x}
\sigma^{\mu\nu}+\sigma^{\mu\nu} \!\not\!{x})  +\cdots \, ,\nonumber\\
S_Q^{ij}(x)&=&\frac{i}{(2\pi)^4}\int d^4k e^{-ik \cdot x} \left\{
\frac{\delta_{ij}}{\!\not\!{k}-m_Q}
-\frac{g_sG^{\alpha\beta}_{ij}}{4}\frac{\sigma_{\alpha\beta}(\!\not\!{k}+m_Q)+(\!\not\!{k}+m_Q)
\sigma_{\alpha\beta}}{(k^2-m_Q^2)^2}\right.\nonumber\\
&&\left.+\frac{\pi^2}{3} \langle \frac{\alpha_sGG}{\pi}\rangle
\delta_{ij}m_Q \frac{k^2+m_Q\!\not\!{k}}{(k^2-m_Q^2)^4}
+\cdots\right\} \, ,
\end{eqnarray}
where $\langle \bar{s}g_s\sigma Gs\rangle=\langle
\bar{s}g_s\sigma_{\alpha\beta} G^{\alpha\beta}s\rangle$  and
$\langle \frac{\alpha_sGG}{\pi}\rangle=\langle
\frac{\alpha_sG_{\alpha\beta}G^{\alpha\beta}}{\pi}\rangle$, then
contract the quark fields in the correlation functions $\Pi^a(p)$
with Wick theorem, and obtain the result:
\begin{eqnarray}
\Pi^a(p)&=&-2i \epsilon^{ijk}\epsilon^{i'j'k'} \int d^4x \, e^{i p
\cdot x} Tr\left[ \gamma_\mu S_{ii'}(x)\gamma_\nu C
S^T_{jj'}(x)C\right]\gamma_5 \gamma_\mu S_Q^{kk'}(x)\gamma_\nu
\gamma_5\, .
\end{eqnarray}
Substitute the full $s$, $c$ and $b$ quark propagators into above
correlation functions and complete  the integral in the  coordinate
space, then integrate over the variable $k$, we can obtain the
correlation functions $\Pi^a_i(p)$ at the level of quark-gluon
degree of freedom. Once  analytical results are obtained,
  then we can take  the quark-hadron duality and perform  Borel transform with respect to the variable
$P^2=-p^2$, finally we obtain  the following two sum rules with
respect to the tensor structures  $\!\not\!{p}$ and $1$
respectively:
\begin{eqnarray}
\lambda^2_{a}\exp(-\frac{M_{\Omega_a}^2}{M^2})&=& \frac{1}{16\pi^4}
\int_{th}^{s^0_{a}}ds \int_{\Delta^a}^1 dx x(1-x)^3\left(
\widetilde{m}_a^2-s\right)\left( 3\widetilde{m}_a^2-5s\right)
\exp(-\frac{s}{M^2})
\nonumber \\
&&+\frac{m_s\langle
\bar{s}s\rangle}{\pi^2}\int_{th}^{s^0_{a}}ds\int_{\Delta^a}^1 dx
x(2-3x)\exp(-\frac{s}{M^2}) \nonumber \\
&&+\frac{1}{48\pi^2}\langle \frac{\alpha_sGG}{\pi}\rangle
\int_{th}^{s^0_{a}}ds\int_{\Delta^a}^1
dx  (4-5x) \exp(-\frac{s}{M^2})\nonumber\\
&&+\frac{m_s\langle \bar{s}s\rangle}{\pi^2} \int_0^1 dx
x(1-x)\widetilde{m}_a^2\exp(-\frac{\widetilde{m}_a^2}{M^2})\nonumber\\
&&-\frac{m_s\langle \bar{s}g_s\sigma Gs\rangle}{6\pi^2} \int_0^1 dx
x(2+\frac{\widetilde{m}_a^2}{M^2})\exp(-\frac{\widetilde{m}_a^2}{M^2})\nonumber\\
&&+\frac{1}{144\pi^2}\langle \frac{\alpha_sGG}{\pi}\rangle \int_0^1
dx
(\frac{2x^3-9x^2+9x-2}{x}-\frac{(1-x)^3\widetilde{m}_a^2}{xM^2})\nonumber\\
&&\widetilde{m}_a^2\exp(-\frac{\widetilde{m}_a^2}{M^2}) \nonumber\\
&&+\frac{2\langle \bar{s}s\rangle^2}{3}
\exp(-\frac{m_a^2}{M^2})+\frac{m_s\langle \bar{s}g_s\sigma
Gs\rangle}{4\pi^2} \exp(-\frac{m_a^2}{M^2}) \, \, ,
\end{eqnarray}

\begin{eqnarray}
M_{\Omega_a}\lambda^2_{a}\exp(-\frac{M_{\Omega_a}^2}{M^2})
 &=& \frac{3m_a}{32\pi^4} \int_{th}^{s_a^0}
ds\int_{\Delta^a}^1 dx (1-x)^2\left(
\widetilde{m}_a^2-s\right)^2\exp(-\frac{s}{M^2})
\nonumber \\
&&-\frac{3m_am_s\langle \bar{s}s\rangle}{2\pi^2}\int_{th}^{s_a^0} ds
\int_{\Delta^a}^1 dx \exp(-\frac{s}{M^2}) \nonumber \\
&&+\frac{m_a}{96\pi^2}\langle \frac{\alpha_sGG}{\pi}\rangle
\int_{th}^{s_a^0} ds \int_{\Delta^a}^1
dx  (\frac{2}{x^2}-2x-3) \exp(-\frac{s}{M^2}) \nonumber\\
&&-\frac{m_a}{96\pi^2}\langle \frac{\alpha_sGG}{\pi}\rangle \int_0^1
dx \frac{(1-x)^2}{x} \widetilde{m}_a^2 \exp(-\frac{\widetilde{m}_a^2}{M^2})\nonumber\\
&&+\frac{5m_am_s\langle \bar{s}g_s\sigma Gs\rangle}{12\pi^2}
\exp(-\frac{m^2_a}{M^2}) +\frac{4m_a\langle \bar{s}s\rangle^2}{3}
\exp(-\frac{m^2_a}{M^2})\, \, , \nonumber\\
\end{eqnarray}
where $th=(m_a+2m_s)^2$,  $\Delta^a=\frac{m_a^2}{s}$ and
$\widetilde{m}_a^2=\frac{m_a^2}{x}$.

  We carry out the  operator
product expansion to the vacuum condensates adding up to dimension-6
and calculate the same Feynman diagrams as in the sum rules for the
$\frac{3}{2}^+$ heavy  baryons \cite{Wang0704} (where the two tensor
structures $g_{\mu\nu}\!\not\!{p}$ and $g_{\mu\nu}$ are chosen), the
contribution from each of the Feynman diagrams differs from the
corresponding one in our previous work.

In calculation, we
 take  assumption of vacuum saturation for the high
dimension vacuum condensates, they  are always
 factorized to lower condensates with vacuum saturation in the QCD sum rules,
  factorization works well in  large $N_c$ limit.
In this article, we take into account the contributions from the
quark condensates $\langle \bar{s}s \rangle$, $\langle \bar{s}s
\rangle^2$,  mixed condensate $\langle \bar{s}g_s \sigma  G{s}
\rangle $, gluon condensate $\langle \frac{\alpha_s
GG}{\pi}\rangle$, and neglect the contributions  from other high
dimension condensates, which are suppressed by large denominators
and would not play significant roles.

Differentiate the above sum rules with respect to the variable
$\frac{1}{M^2}$, then eliminate the pole residue
$\lambda_{\Omega_a}$, we obtain two QCD sum rules for the masses
$M_{\Omega_a}$ with respect to the tensor structures  $\!\not\!{p}$
and $1$, respectively.

In the heavy quark limit $m_Q\rightarrow\infty$, the quarks $c$ and
$b$ degenerate, we replace the heavy quark field $Q_a(x)$ in the
 baryon current $J^a(x)$ with the effective  field $h(x)$ to
obtain the interpolating current
$J_h(x)=\epsilon^{ijk}s^T_i(x)C\gamma_\mu s_j(x)\gamma_5 \gamma^\mu
h_k(x)$ in the heavy quark effective theory, which  differs  from
the corresponding one constructed from the heavy quark effective
theory remarkably and does not  warrant the tensor structure
$\frac{1+\!\not\!{v}}{2}$ \cite{Grozin92}.

 In the heavy
quark limit, the correlation function $\Pi^a(\omega)$ (with
$\omega=v \cdot p$) can be written as
$\Pi^a(\omega)=\!\not\!{v}\Pi_1(\omega) +\Pi_2(\omega)
$($\Pi_1(\omega) \neq\Pi_2(\omega)$). In Ref.\cite{Liu07}, the
current operator $\eta(x)=\epsilon^{ijk}s^T_i(x)C\gamma_\mu
s_j(x)\gamma_5 [\gamma^\mu-\!\not\!{v}v^\mu] h_k(x)$ is used to
interpolate the baryons  $\Omega_a$, and the correlation function is
written as $\Pi^a(\omega)=\frac{1+\!\not\!{v}}{2}\bar{\Pi}(\omega)$.

The effective heavy quark propagator has the following form,
\begin{eqnarray}
S_h^{ab}(x)&=&\delta_{ab}\frac{1+\!\not\!{v}}{2} \int_0^\infty dt
\delta(x-vt) \, ,
\end{eqnarray}
where the  $v_\mu$ is a four-vector with $v^2=1$. The calculations
of the operator product expansion can be performed in the coordinate
space and are greatly facilitated, as we do not need the mixed
picture both in coordinate and momentum spaces. We calculate the
same Feynman diagrams as in the full QCD, the contributions of some
diagrams vanish  in the heavy quark limit.

 Finally we obtain two sum rules with respect to the tensor structures  $\!\not\!{v}$ and
$1$ respectively:
\begin{eqnarray}
F^2\exp(-\frac{\bar{\Lambda}}{T})&=&\frac{1}{5\pi^4}
\int_{2m_s}^{\omega_0}d\omega \omega^5 \exp(-\frac{\omega}{T})
 +\frac{m_s\langle \bar{s}g_s \sigma G s\rangle}{24\pi^2}
+\frac{\langle \bar{s}s\rangle^2}{3}  \, , \\
F^2\exp(-\frac{\bar{\Lambda}}{T})&=& \frac{1}{10\pi^4}
\int_{2m_s}^{\omega_0}d\omega \omega^5 \exp(-\frac{\omega}{T})
-\frac{3m_s\langle
\bar{s}s\rangle}{\pi^2}\int_{2m_s}^{\omega_0}d\omega \omega
\exp(-\frac{\omega}{T})
\nonumber \\
 &&-\frac{1}{16\pi^2}\langle\frac{\alpha_sGG}{\pi}\rangle
\int_{2m_s}^{\omega_0}d\omega \omega
\exp(-\frac{\omega}{T})+\frac{5m_s\langle \bar{s}g_s \sigma G
s\rangle}{24\pi^2} +\frac{2\langle \bar{s}s\rangle^2}{3}  \, ,
\nonumber\\
\end{eqnarray}
where  we have used the definitions for the binding energy
$\bar{\Lambda}$ and the pole residue $F$,
$limit_{m_Q\rightarrow\infty} M_\Omega=m_Q+\bar{\Lambda}$  and
$\langle 0|J(0)|\Omega(v,s)\rangle=F N(v,s)$, the $N(v,s)$ is the
Dirac spin vector in the heavy quark limit.

Differentiate the above sum rules with respect to the variable
$\frac{1}{T}$, then eliminate the pole residue $F$, we obtain two
QCD sum rules for the binding energy $\bar{\Lambda}$ with respect to
the tensor structures  $\!\not\!{v}$ and $1$, respectively.

\section{Numerical results and discussions}
The input parameters are taken to be the standard values $\langle
\bar{q}q \rangle=-(0.24\pm 0.01 \,\rm{GeV})^3$, $\langle \bar{s}s
\rangle=(0.8\pm 0.2 )\langle \bar{q}q \rangle$, $\langle
\bar{s}g_s\sigma Gs \rangle=m_0^2\langle \bar{s}s \rangle$,
$m_0^2=(0.8 \pm 0.2)\,\rm{GeV}^2$, $\langle \frac{\alpha_s
GG}{\pi}\rangle=(0.33\,\rm{GeV})^4 $, $m_s=(0.14\pm0.01)\,\rm{GeV}$,
$m_c=(1.4\pm0.1)\,\rm{GeV}$ and $m_b=(4.8\pm0.1)\,\rm{GeV}$ at the
energy scale about $\mu=1\, \rm{GeV}$ \cite{SVZ79,PRT85,Ioffe2005}.
 The contribution from the gluon condensate $\langle
\frac{\alpha_s GG}{\pi}\rangle $ is less than $3\%$, and the
uncertainty is neglected here.

 We usually  consult the
experimental data in choosing the Borel parameter $M^2$ and the
threshold parameter $s_0$. There lack  experimental data for the
phenomenological hadronic spectral densities of the bottomed baryons
at present, only the $\Lambda_b$, $\Sigma_b$, $\Sigma_b^*$, $\Xi_b$
\cite{PDG} and $\Omega_b^-$ \cite{OmegaB} have been  observed, we
can borrow some ideas from the light baryon spectra \cite{PDG}.

For the octet baryons with $J^{P}={\frac{1}{2}}^+$, the mass of the
proton   is $M_p=938\,\rm{MeV}$, and the mass of the first radial
excited state $N(1440)$ (the Roper resonance) is
$M_{1440}=(1420-1470)\,\rm{MeV}\approx 1440\,\rm{MeV}$;
 the mass of the ground
state $\Sigma$ is  $M_{\Sigma}=1189.37 \, \rm{MeV}$, and the mass of
the first radial excited state $\Sigma(1660)$ is
$M_{1660}=(1630-1690) \, \rm{MeV}\approx 1660 \, \rm{MeV}$; the mass
of the ground state $\Xi$  is  $M_{\Xi}=1321.7\, \rm{MeV}$, while
the spin-parity of the high excited states $\Xi(1620)$, $\Xi(1690)$,
$\Xi(1950)$, $\Xi(2030)$, $\Xi(2120)$, $\Xi(2250)$, $\Xi(2370)$,
$\Xi(2500)$ have not been determined yet \cite{PDG}. For the
decuplet  baryons with $J^{P}={\frac{3}{2}}^+$, the mass of  the
$\Delta(1232)$  is $M_{1232}=(1231-1233)\,\rm{MeV}\approx
1232\,\rm{MeV}$,  and the mass of the first radial excited state
$\Delta(1600)$ is $M_{1600}=(1550-1700)\,\rm{MeV}\approx
1600\,\rm{MeV}$;  the mass of the ground state $\Sigma(1385)$ is
$M_{1385}=1382.8 \, \rm{MeV}$, and the mass of the first radial
excited state $\Sigma(1840)$ is $M_{1840}= 1840 \, \rm{MeV}$; the
ground states $\Xi(1530)$, $\Omega(1672)$ are well established,
while the spin-parity of the high excited states $\Omega(2250)$,
$\Omega(2380)$, $\Omega(2470)$ have not been determined yet
\cite{PDG}.

 From the  experimental data for the baryons consist of   $qqq$ and $qqs$,
 we can see that the energy gap between
 the ground states and the first radial excited states
is about $(0.4-0.5)\,\rm{GeV}$, the $SU(3)$ breaking effects are
rather small. So in the QCD sum rules for the baryons with the light
quarks, the threshold parameters $s_0$ are always chosen to be
$\sqrt{s_0}=M_{gr}+0.5\,\rm{GeV}$
\cite{Ioffe81,Ioffe84,Ioffe82,Ioffe83}, here $gr$ stands for the
ground states. The central values of the threshold parameters for
the heavy baryons $\Omega_c^0$ and $\Omega_b^-$ can be chosen as
$s^0_{\Omega_c}=(2.6975+0.5)^2 \,\rm{GeV}^2$ and $s^0_{\Omega_b} =
(6.165+0.5)^2 \,\rm{GeV}^2$, respectively.

\begin{table}
\begin{center}
\begin{tabular}{|c|c|c|}
\hline\hline & Eq.(8)& Eq.(9)\\ \hline
      $\mbox{perturbative term}$  &$+93\%$ &$+70\%$\\ \hline
      $ \langle \bar{s} s\rangle$& $-1\%$ &$+21\%$\\      \hline
     $ \langle \bar{s} g_s\sigma G s\rangle$& $-1\%$ &$-3\%$\\     \hline
    $\langle\bar{s} s\rangle^2$&  $+8\%$ &$+10\%$\\ \hline
$\langle\frac{\alpha_s GG}{\pi}\rangle$&  $+1\%$ &$+1\%$\\ \hline
    \hline
\end{tabular}
\end{center}
\caption{ The contributions from different terms  in the sum rules
for the $\Omega_c^0$ with the central values of the input parameters
"Set I". }
\end{table}

\begin{table}
\begin{center}
\begin{tabular}{|c|c|c|}
\hline\hline & Eq.(8)& Eq.(9)\\ \hline
      $\mbox{perturbative term}$  &$+94\%$ &$+76\%$\\ \hline
      $ \langle \bar{s} s\rangle$& $-2\%$ &$+23\%$\\      \hline
     $ \langle \bar{s} g_s\sigma G s\rangle$& $-1\%$ &$-2\%$\\     \hline
    $\langle\bar{s} s\rangle^2$&  $+8\%$ &$+5\%$\\ \hline
$\langle\frac{\alpha_s GG}{\pi}\rangle$&  $+1\%$ &$-3\%$\\ \hline
    \hline
\end{tabular}
\end{center}
\caption{ The contributions from different terms  in the sum rules
for the $\Omega_b^-$ with the central values of the input parameters
"Set I".}
\end{table}

In this article, the threshold parameters and the Borel parameters
are taken as $s^0_{\Omega_c}=(10.5\pm1.0)\,\rm{GeV}^2$ and
$M^2=(2.2-3.2)\,\rm{GeV}^2$ for the charmed  baryon  $\Omega_c^0$,
and $s^0_{\Omega_b}=(44.5\pm1.0)\,\rm{GeV}^2$ and
$M^2=(5.0-6.0)\,\rm{GeV}^2$ for the bottomed baryon $\Omega_b^-$;
thereafter these parameters will be denoted as "Set I". In the heavy
quark limit $m_Q\rightarrow\infty$, the corresponding Borel
parameter $T$ and threshold parameter $\omega_0$ can be taken as
$T=(0.4-0.6)\, \rm{GeV}$ and $\omega_0=(1.7-1.9)\, \rm{GeV}$; and we
 will refer them  as "Set III".

 We can make another estimation for the threshold parameters by
adding the heavy quark masses to the ground state $\frac{1}{2}^+$
baryons which consist of $qss$,
$s^0_{\Omega_c}=(M_\Xi+m_c+0.5\,\rm{GeV})^2\approx 10.5\,\rm{GeV}^2$
and $s^0_{\Omega_b}=(M_\Xi+m_b+0.5\,\rm{GeV})^2\approx
44.0\,\rm{GeV}^2$, which are almost  the same values as "Set I".

The contributions from different terms at the central values of the
input parameters are presented in Table.1 and Table.2, respectively.
From the two tables, we can expect convergence of the operator
product expansion.  The components $\Pi^a_1(p)$ with the odd
structure have smaller dimension of mass than  the components
$\Pi^a_2(p)$ with the even structure, see Eq.(4), it is not
unexpected that the components $\Pi^a_1(p)$ have better convergent
behavior in the operator product expansion and result in better QCD
sum rules, see Table 4.

In the heavy quark limit, the contribution from the quark condensate
$\langle \bar{s}s\rangle$ vanishes for the odd structure
$\!\not\!{v}$, while the  condensate $\langle \bar{s}s\rangle$ has a
numerical coefficient $\frac{3}{\pi^2}$ for the even structure $1$,
see Eqs.(11-12), which can explain the hierarchy appears in the
operator product expansion naturally, see Tables 1-2.

The contributions of the pole terms for different sum rules are
shown in Table 3, from the table, we can see that the pole dominance
condition is marginally satisfied for the parameters "Set I".

We can choose smaller Borel parameters  $M^2$ or larger threshold
parameters $s^0_a$ to enhance the contributions from the ground
states. However, if we take larger threshold parameter $s^0_a$, the
contribution from the first radial excited state maybe included in;
on the other hand, for smaller Borel parameter $M^2$, the sum rules
are not stable enough, the uncertainty with variation of the Borel
parameter is large.

In this article, we also present the results with smaller Borel
parameters  $M^2$ and larger threshold parameters $s^0_a$,
$s^0_{\Omega_c}=(11.0\pm1.0)\,\rm{GeV}^2$ and
$M^2=(1.9-2.6)\,\rm{GeV}^2$ for the charmed  baryon $\Omega_c^0$,
and $s^0_{\Omega_b}=(46.5\pm1.0)\,\rm{GeV}^2$ and
$M^2=(4.5-5.4)\,\rm{GeV}^2$ for the bottomed baryon $\Omega_b^-$.
These parameters (denoted as "Set II")  satisfy both  pole dominance
(see Table 3) and convergence of the operator product expansion.

\begin{figure}
 \centering
 \includegraphics[totalheight=6cm,width=7cm]{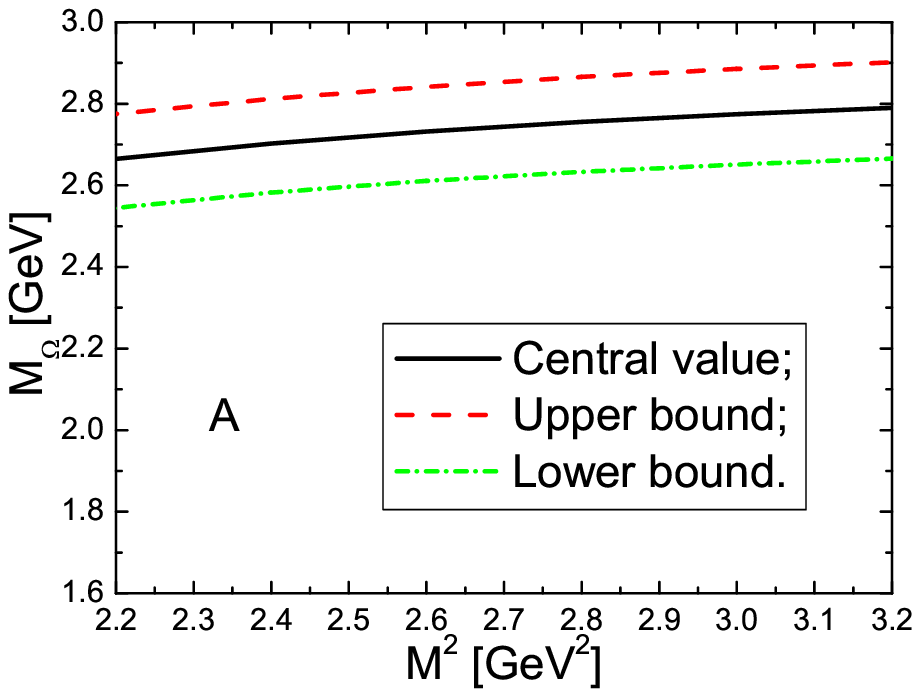}
  \includegraphics[totalheight=6cm,width=7cm]{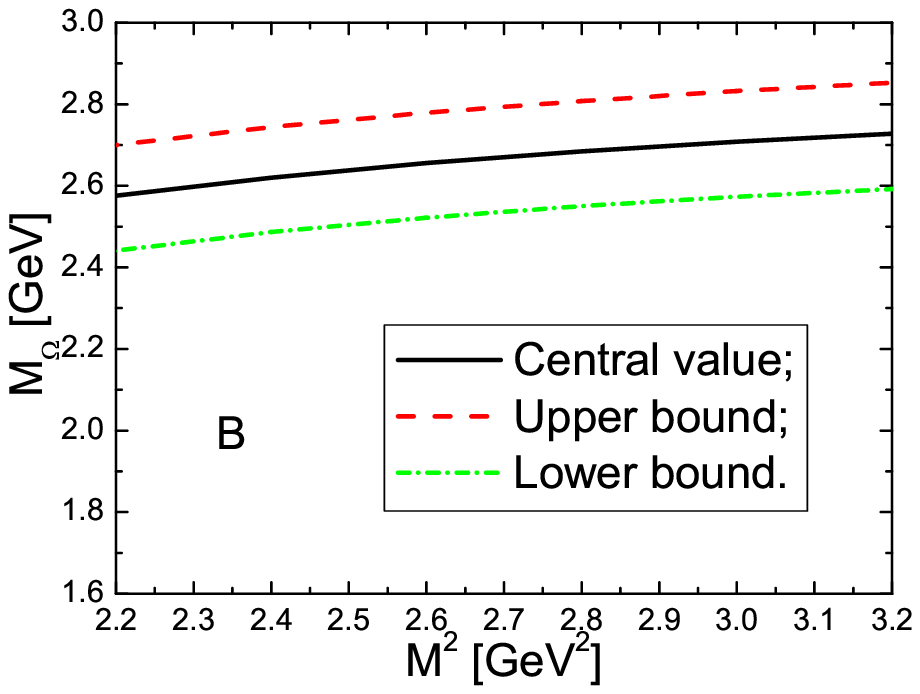}
  \includegraphics[totalheight=6cm,width=7cm]{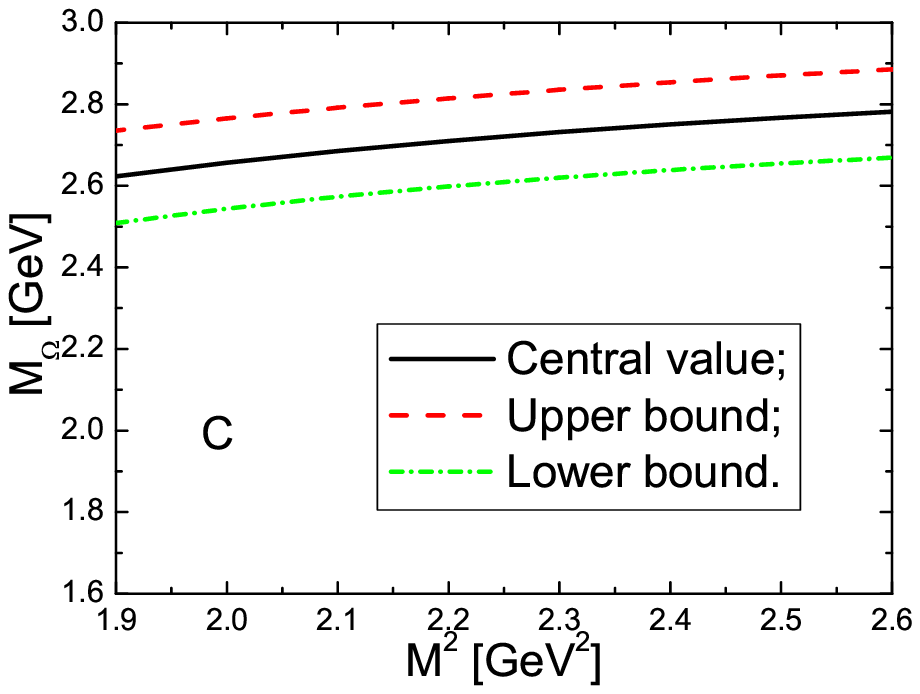}
  \includegraphics[totalheight=6cm,width=7cm]{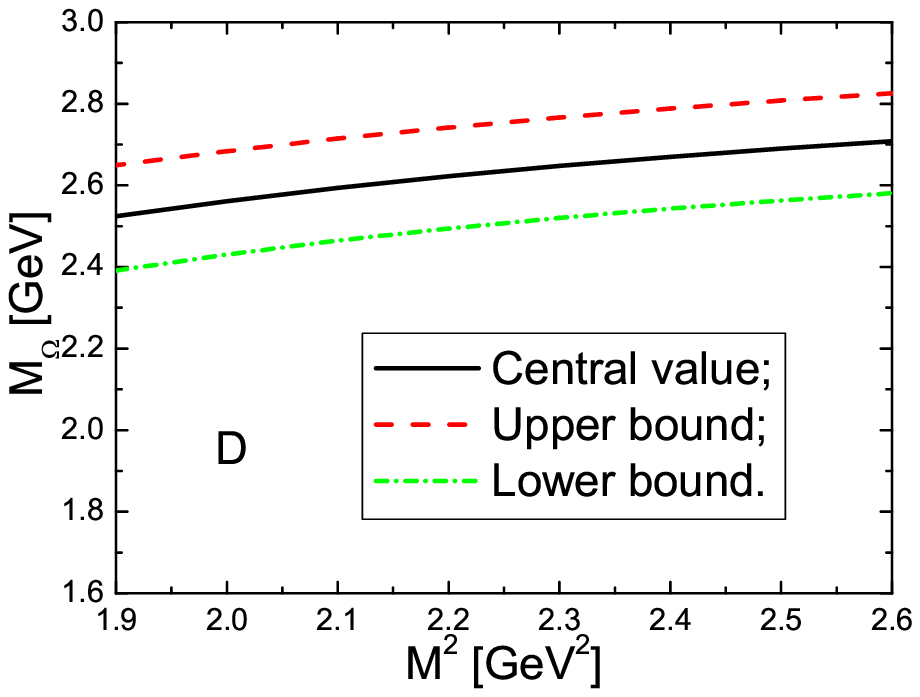}
  \caption{ The mass $M_{\Omega_c^0}$ from the sum rules   with
different tensor structures and input parameters, $A$, $B$, $C$ and
$D$ correspond to   $\!\not\!{p} \,\,\&\,\, \rm{Set \, I }$, $ 1
\,\,\&\,\, \rm{Set \, I }$, $ \!\not\!{p} \,\,\&\,\, \rm{Set \, II
}$ and $1 \,\,\&\,\, \rm{Set \, II }$, respectively. }
\end{figure}

\begin{figure}
 \centering
 \includegraphics[totalheight=6cm,width=7cm]{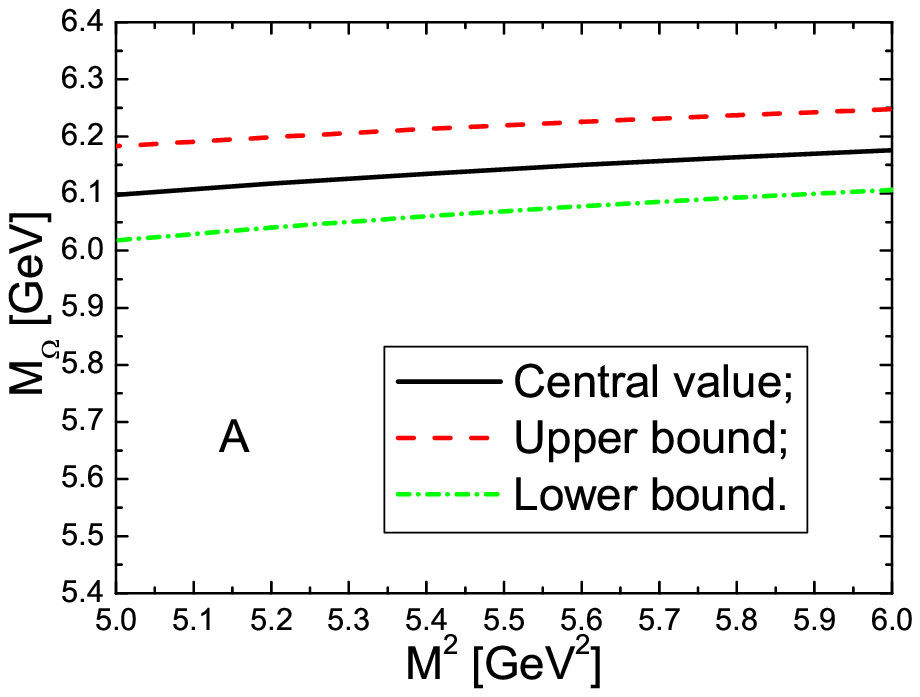}
  \includegraphics[totalheight=6cm,width=7cm]{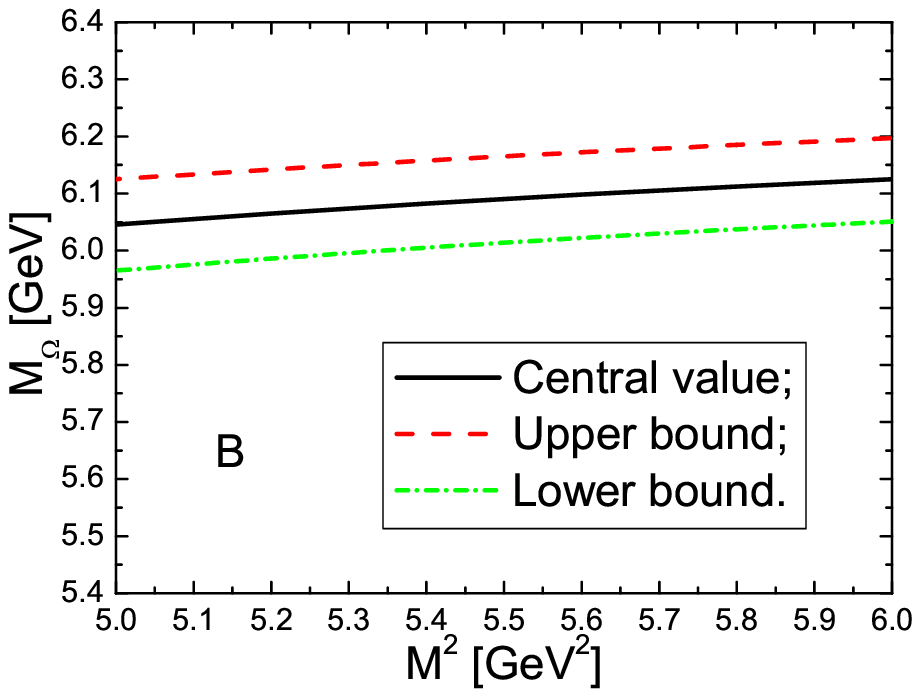}
  \includegraphics[totalheight=6cm,width=7cm]{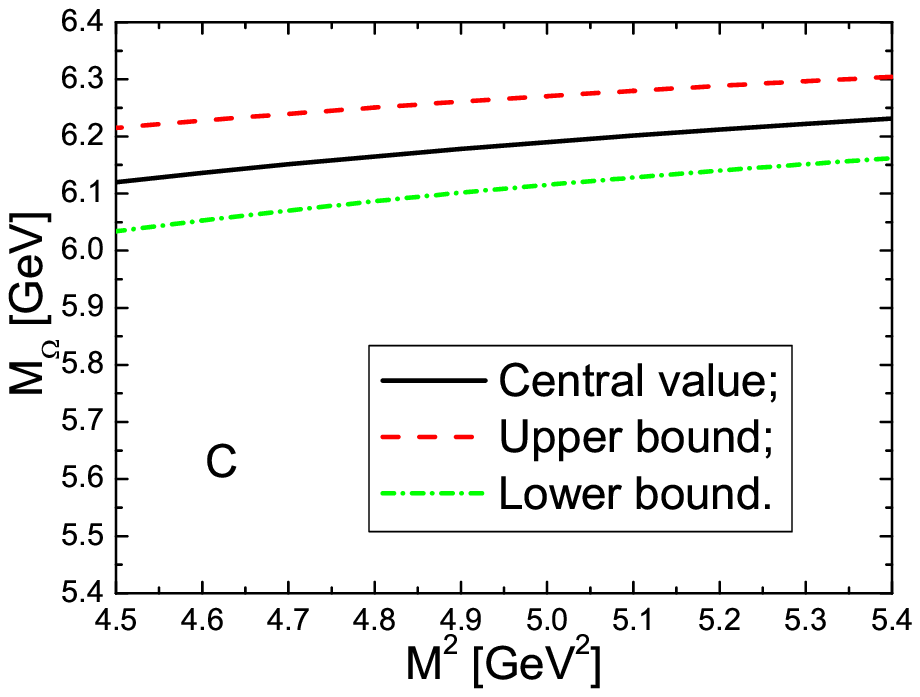}
  \includegraphics[totalheight=6cm,width=7cm]{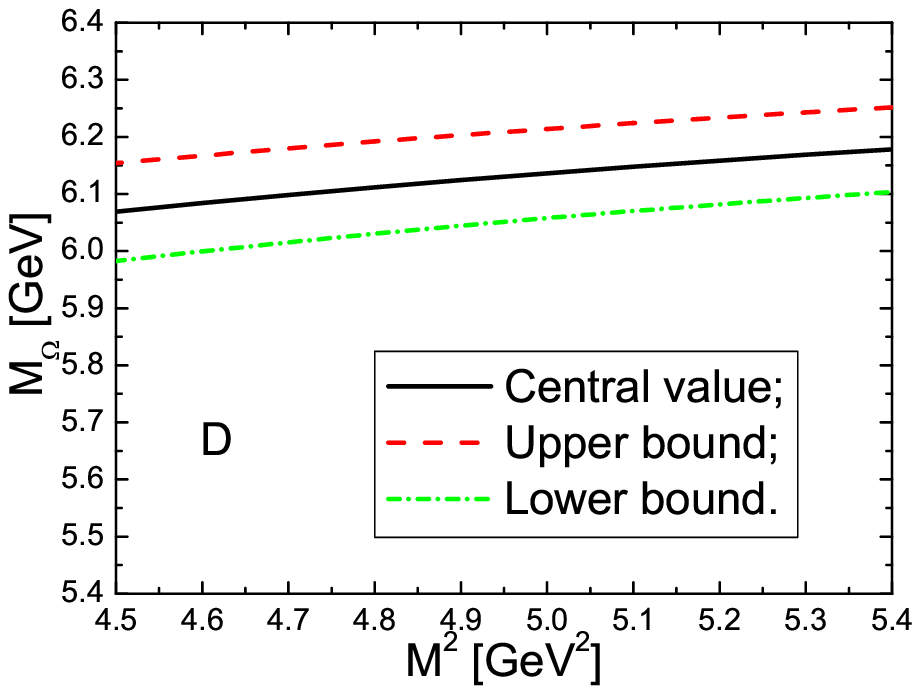}
  \caption{ The mass $M_{\Omega_b^-}$ from the sum rules   with
different tensor structures and input parameters, $A$, $B$, $C$ and
$D$ correspond to   $\!\not\!{p} \,\,\&\,\, \rm{Set \, I }$, $ 1
\,\,\&\,\, \rm{Set \, I }$, $ \!\not\!{p} \,\,\&\,\, \rm{Set \, II
}$ and $1 \,\,\&\,\, \rm{Set \, II }$, respectively. }
\end{figure}

\begin{figure}
 \centering
 \includegraphics[totalheight=6cm,width=7cm]{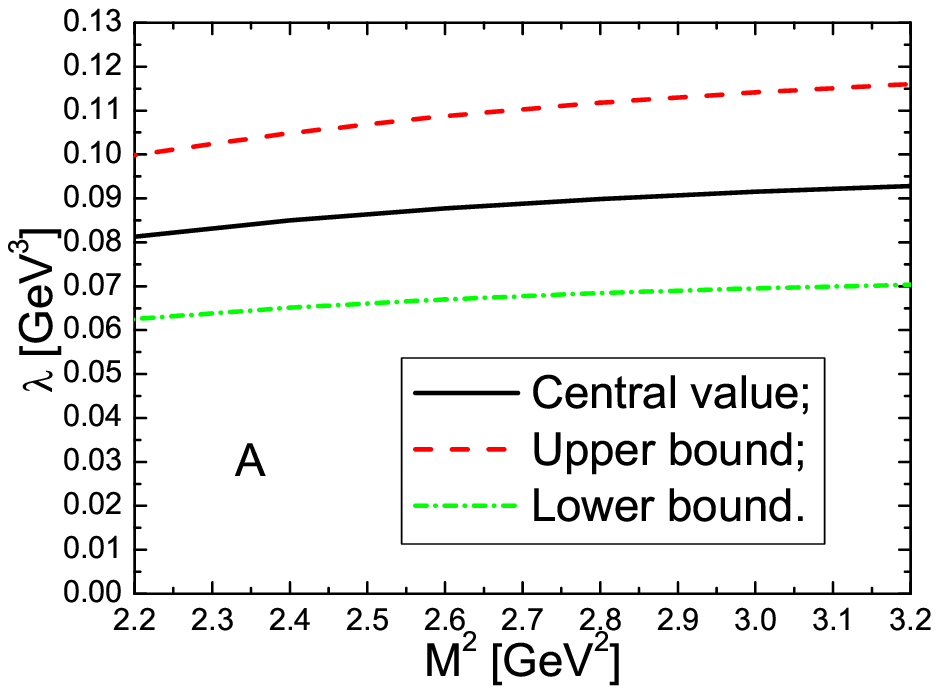}
 \includegraphics[totalheight=6cm,width=7cm]{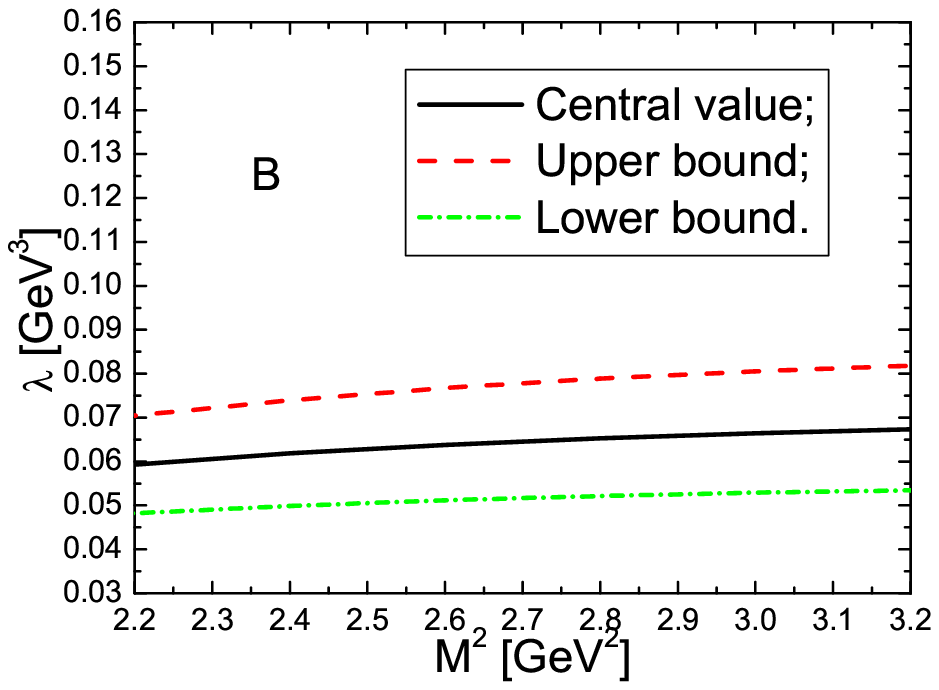}
 \includegraphics[totalheight=6cm,width=7cm]{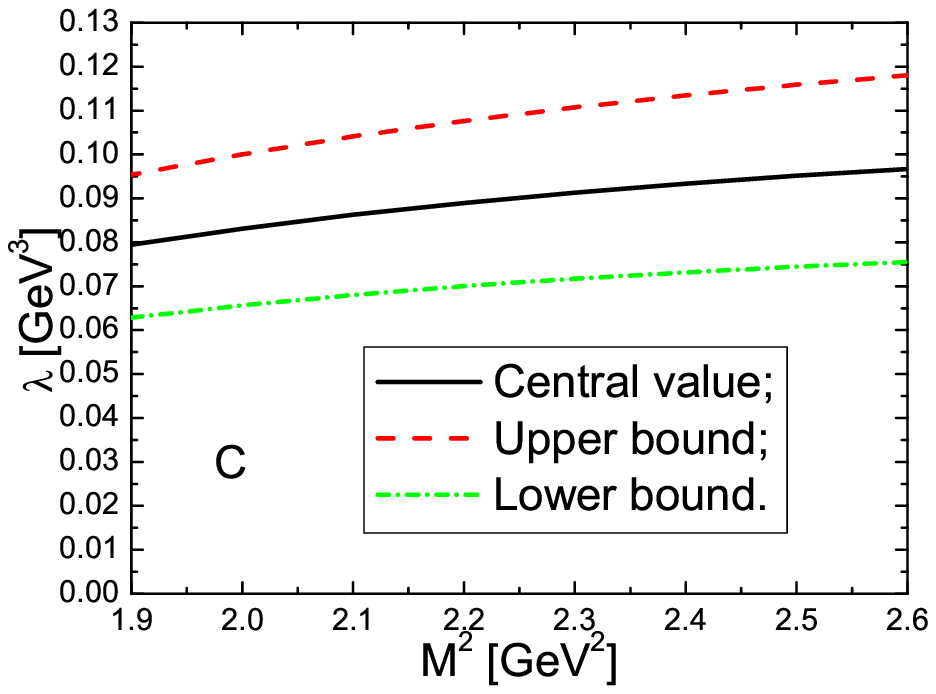}
  \includegraphics[totalheight=6cm,width=7cm]{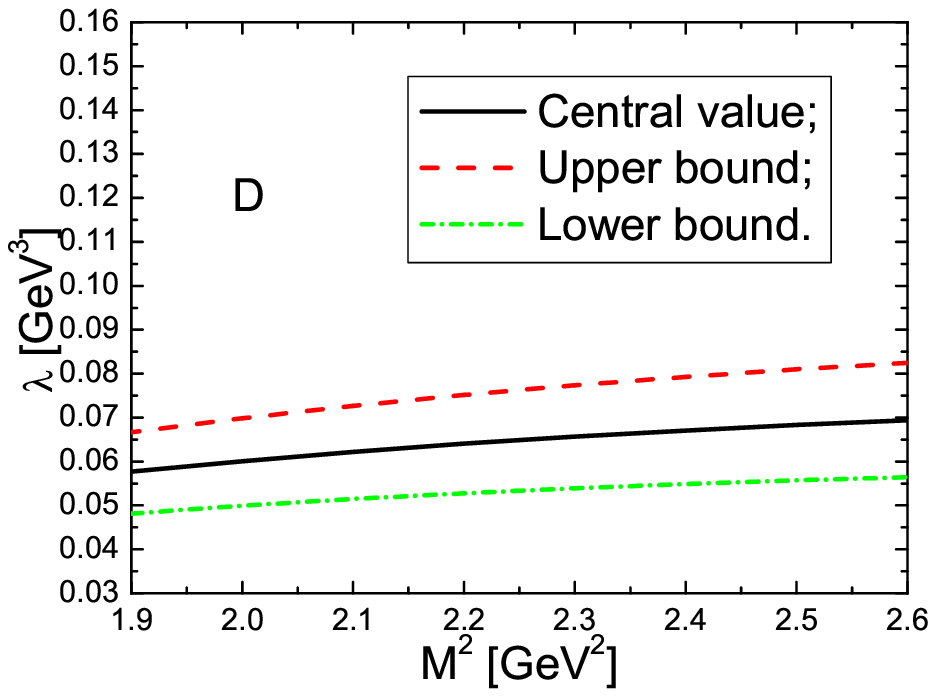}
  \caption{The pole residue $\lambda_{\Omega_c}$ from the sum rules   with
different tensor structures and input parameters, $A$, $B$, $C$ and
$D$ correspond to   $\!\not\!{p} \,\,\&\,\, \rm{Set \, I }$, $ 1
\,\,\&\,\, \rm{Set \, I }$, $ \!\not\!{p} \,\,\&\,\, \rm{Set \, II
}$ and $1 \,\,\&\,\, \rm{Set \, II }$, respectively. }
\end{figure}

\begin{figure}
 \centering
 \includegraphics[totalheight=6cm,width=7cm]{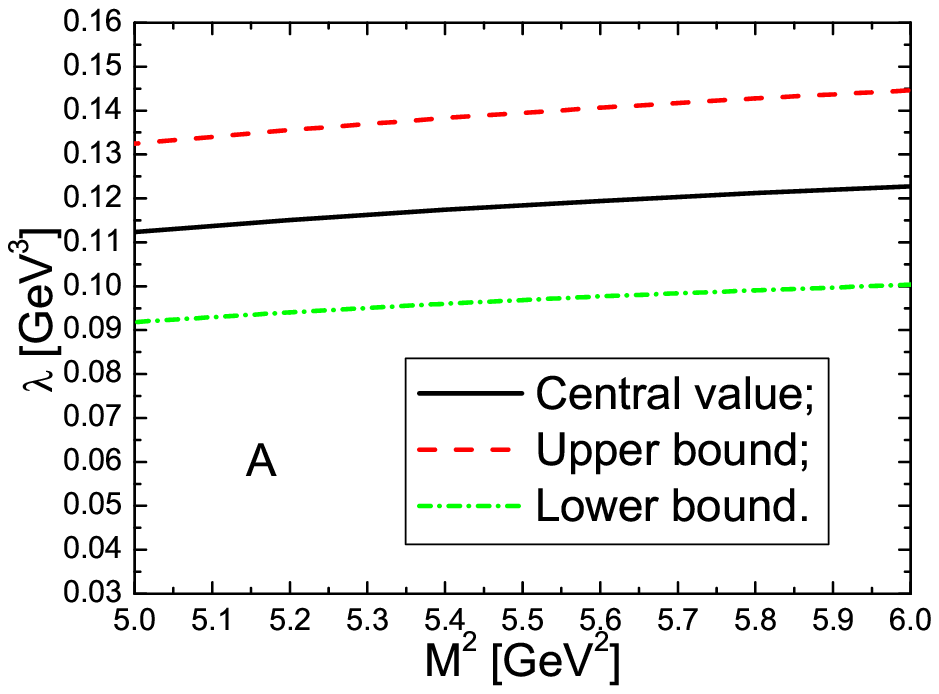}
 \includegraphics[totalheight=6cm,width=7cm]{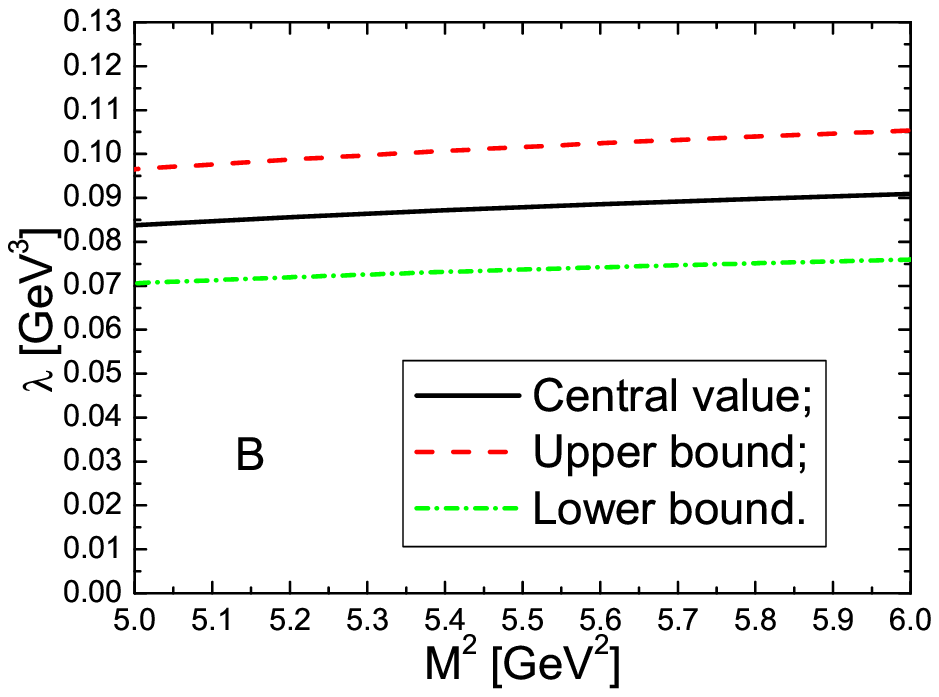}
 \includegraphics[totalheight=6cm,width=7cm]{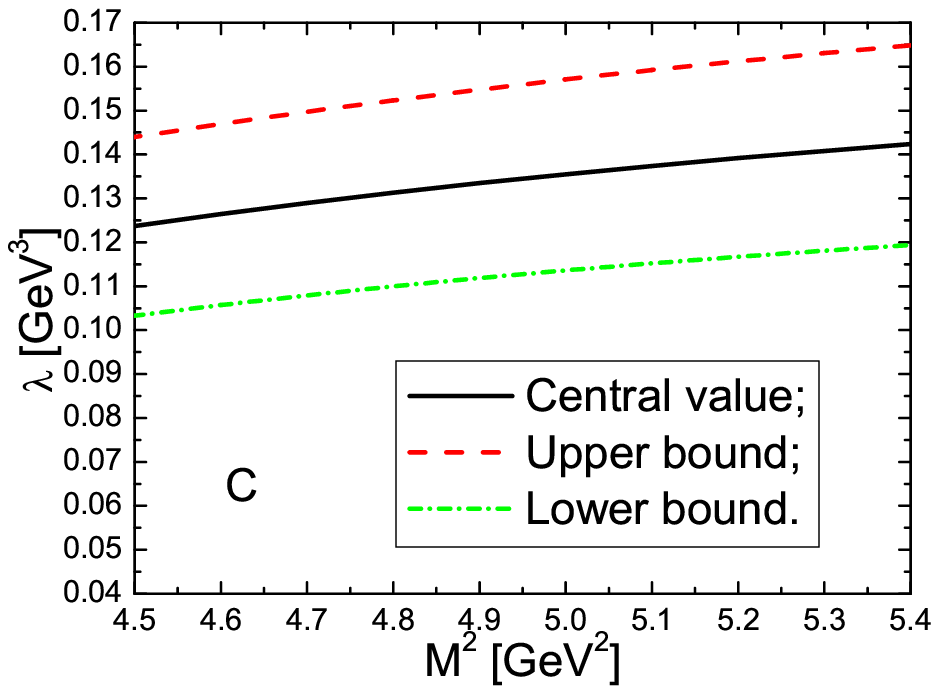}
 \includegraphics[totalheight=6cm,width=7cm]{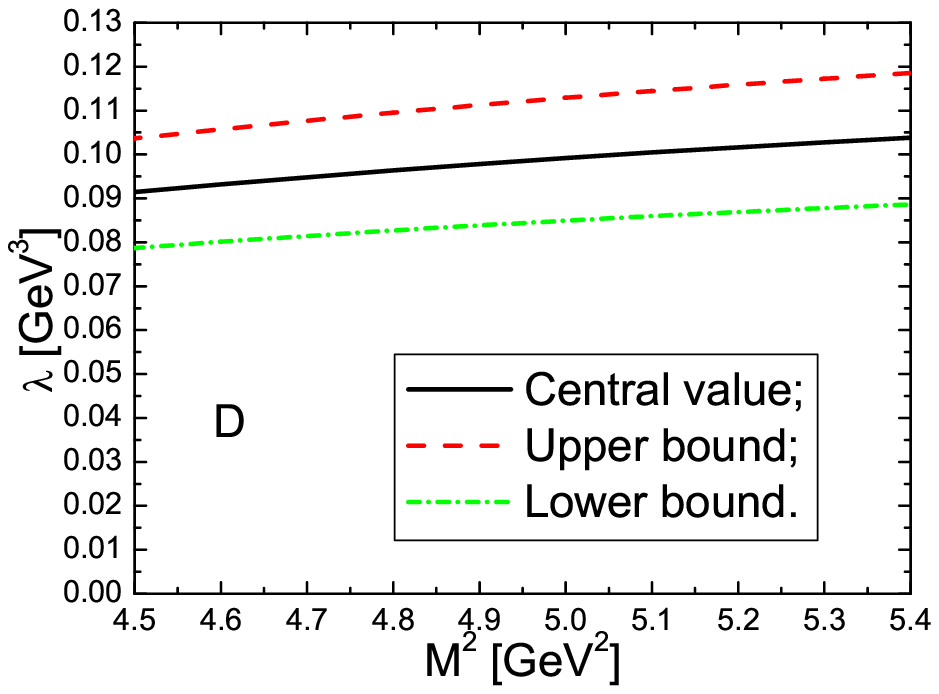}
  \caption{The pole residue $\lambda_{\Omega_b}$ from the sum rules   with
different tensor structures and input parameters, $A$, $B$, $C$ and
$D$ correspond to   $\!\not\!{p} \,\,\&\,\, \rm{Set \, I }$, $ 1
\,\,\&\,\, \rm{Set \, I }$, $ \!\not\!{p} \,\,\&\,\, \rm{Set \, II
}$ and $1 \,\,\&\,\, \rm{Set \, II }$, respectively. }
\end{figure}

\begin{figure}
 \centering
 \includegraphics[totalheight=6cm,width=7cm]{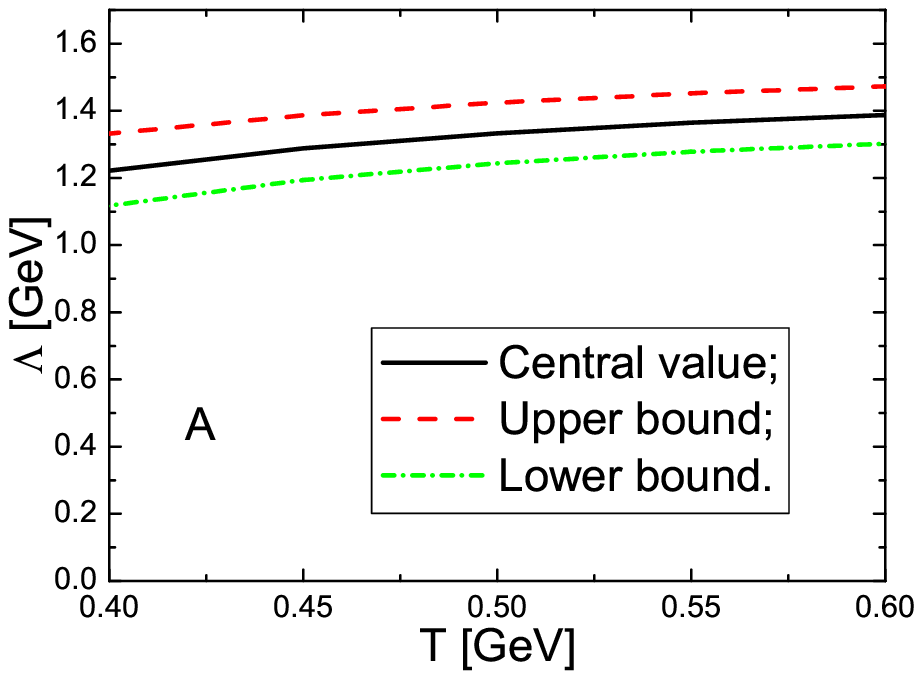}
  \includegraphics[totalheight=6cm,width=7cm]{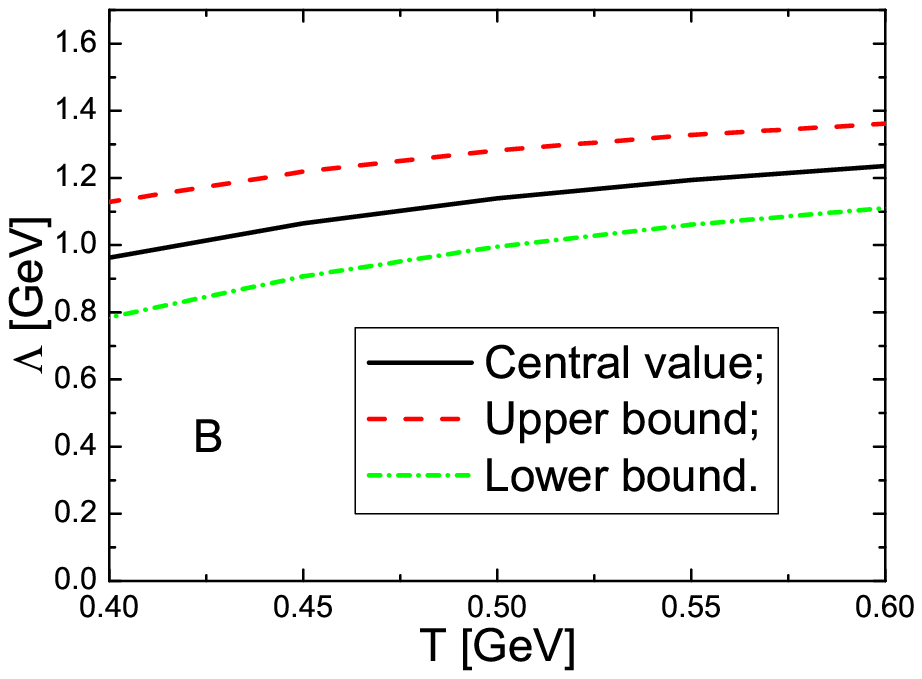}
  \caption{ The binding energy $\bar{\Lambda}$ with variation of the  Borel parameter $T$, $A$ and $B$ correspond to the tensor
  structures $ \!\not\!{v} $ and $1 $, respectively.}
\end{figure}

Taking into account all uncertainties of the input parameters,
finally we obtain the values of the masses and the pole residues of
 the heavy baryons $\Omega_c^0$ and $\Omega_b^-$, which are
shown in Figs.1-4 (and Table 4), respectively.

In the heavy quark limit, we  obtain the binding energy
$\bar{\Lambda}$ (see Fig.5)
\begin{eqnarray}
\bar{\Lambda} &=& 1.30\pm 0.17\,\rm{GeV} \, , \nonumber\\
\bar{\Lambda} &=& 1.10\pm 0.26\,\rm{GeV} \, ,
\end{eqnarray}
for the tensor structures  $\!\not\!{v}$ and $1$ respectively. The
corresponding masses  are presented  in Table 4.

\begin{table}
\begin{center}
\begin{tabular}{|c|c|c|c|}
\hline\hline & $\Omega_c^0$& $\Omega_b^-$\\ \hline
      $\!\not\!{p} \,\,\&\,\, \rm{Set \, I }$  &$(34-60)\%$ &$(35-49)\%$\\ \hline
       $ 1 \,\,\&\,\, \rm{Set \, I }$& $(34-62)\%$& $(36-51)\%$ \\     \hline
      $ \!\not\!{p} \,\,\&\,\, \rm{Set \, II }$ &$(52-75)\%$ &$(51-66)\%$\\      \hline
    $1 \,\,\&\,\, \rm{Set \, II }$&  $(53-78)\%$ &  $(51-67)\%$\\ \hline
    \hline
\end{tabular}
\end{center}
\caption{ The contributions of the  pole terms from different sum
rules, here we take the central values of the input parameters
except for the Borel parameter $M^2$.}
\end{table}

\begin{table}
\begin{center}
\begin{tabular}{|c|c|c|c|c|}
\hline\hline &$M_{\Omega_c^0}(\rm{GeV})$&$M_{\Omega_b^-}(\rm{GeV})$&
$\lambda_{\Omega_c^0}(\rm{GeV}^3)$&$\lambda_{\Omega_b^-}(\rm{GeV}^3)$\\
\hline $\!\not\!{p} \,\,\&\,\, \rm{Set \, I } $&  $2.72\pm0.18$ & $6.13\pm0.12$ & $0.088\pm 0.026$ &$0.118\pm 0.026$ \\
\hline
$1 \,\,\&\,\, \rm{Set \, I } $&$2.66\pm0.19$ & $6.09\pm0.11$&$0.064\pm 0.017$ &$0.088\pm 0.017$ \\
\hline
$\!\not\!{p} \,\,\&\,\, \rm{Set \, II } $&$2.71\pm0.18$ & $6.18\pm0.13$&$0.090\pm0.027$ & $0.133\pm0.030$\\
\hline
$1 \,\,\&\,\, \rm{Set \, II } $& $2.63\pm0.20$& $6.12\pm0.13$& $0.065\pm0.017$& $0.099\pm0.019$\\
\hline
$\!\not\!{v} \,\,\&\,\, \rm{Set \, III } $& $2.70\pm0.27$& $6.10\pm0.27$& & \\
\hline
$1 \,\,\&\,\, \rm{Set \, III } $& $2.50\pm0.36$& $5.90\pm0.36$& & \\
\hline $\rm{Exp \,\, data } $& $2.6975\pm0.0026$& $6.165 \pm
0.013\thinspace $& & \\
\hline
         \hline
\end{tabular}
\end{center}
\caption{ The masses and pole residues from the sum rules with
different tensor structures and input parameters. }
\end{table}

The values  $M_{\Omega_c^0}=(2.72\pm0.18)\,\rm{GeV}$ and
$M_{\Omega_c^0}=(2.71\pm0.18)\,\rm{GeV}$ from the sum rules with the
tensor structure $\!\not\!{p}$ are in good agreement with the
experimental data $M_{\Omega_c^0}=(2.6975\pm0.0026) \,\rm{GeV}$
\cite{PDG}, other theoretical predictions also indicate the value is
about $2.7\, \rm{GeV}$
\cite{Roncaglia95,Valcarce08,Jenkins96,Bowler96,Mathur02,Ebert05,Ebert08,Karliner07,Liu07,Roberts07}.
The experimental value $M_{\Omega_b^-}=6.165\pm 0.010\thinspace \pm
0.013\thinspace \, \rm{GeV}$ is about $0.1 \, \rm{GeV}$ larger than
the existing theoretical calculations
\cite{Roncaglia95,Valcarce08,Jenkins96,Bowler96,Mathur02,Ebert05,Ebert08,Karliner07,Liu07,Roberts07}
(including the QCD sum rules in the heavy quark effective theory
with  $1/m_Q$ corrections \cite{Liu07}), our predictions
$M_{\Omega_b^-}=(6.13\pm0.12)\,\rm{GeV}$  and
$M_{\Omega_b^-}=(6.18\pm0.13)\,\rm{GeV}$  based on the sum rules
with the tensor structure $\!\not\!{p}$ are  excellent.

The parameters "Set I" satisfy the pole dominance criterion
marginally, the parameters "Set II" have larger threshold parameters
than  that of the "Set I", which may take into account some
contributions from the high resonances. More experimental data are
needed  to select the ideal sum rules. Once reasonable values of the
pole residues $\lambda_{\Omega_c}$ and $\lambda_{\Omega_b}$ are
obtained, we can take them as   basic input parameters and study the
hadronic processes with the (light-cone) QCD sum rules.

 In Ref.\cite{Nielsen07}, the tensor structure
$\!\not\!{p}$ is chosen, and the spectral density (including the
perturbative term, quark condensate term $\langle \bar{s}s\rangle$,
mixed condensate term $\langle \bar{s}g_s\sigma Gs \rangle $ and
gluon condensate term $\langle \frac{\alpha_s GG}{\pi}\rangle$)
differs from mine, the most evident difference is that the terms of
the vacuum condensate $\langle \bar{s}s\rangle$ present in Eq.(8)
are absent. I check all calculations carefully and confirm my
results. For example, although the term $\langle \bar{s}s\rangle$
vanishes in the heavy quark limit (see Eq.(11)), there indeed exist
such a term in the full QCD. The threshold parameter
$s^0_{\Omega_c}=(10.0-11.5)\,\rm{GeV}^2$ and the prediction
$M_{\Omega_c^0}=(2.65\pm0.25)\,\rm{GeV}$ are consistent with mine,
while the lower bound of the threshold parameter
$s^0_{\Omega_b}=(41.0-45.0)\,\rm{GeV}^2$ is much smaller than mine,
and the prediction $M_{\Omega_b^-}=(5.82\pm0.23)\,\rm{GeV}$ differs
from the experimental data remarkably.

 The  sum rules in the heavy quark limit with the
odd structure $\!\not\!{v}$ also result in excellent values
$M_{\Omega_c^0}=(2.70\pm0.27)\,\rm{GeV} $ and
 $M_{\Omega_b^-}=(6.10\pm0.27)\,\rm{GeV}$,  although the uncertainties are
 large;  the values from the even structure $1$ are not good enough, see Table 4, the
$1/m_Q$ corrections maybe large. In Ref.\cite{Liu07}, the threshold
parameter is taken as $\omega_0=1.55\, \rm{GeV}$, which differs from
mine remarkably.  The predictions (the central values are
$M_{\Omega_c^0}=2.62\,\rm{GeV} $ and
 $M_{\Omega_b^-}=5.97\,\rm{GeV}$ without the $1/m_Q$ corrections)
 are smaller than mine about $(0.08-0.10)\,\rm{GeV}$. It is not unexpected,   different
interpolating currents and input parameters can lead to different
results.

In this article, we do not take into account the  next-to-leading
order corrections to the perturbative term, the corrections maybe
large. In the massless limit $m_s=m_u=m_d=0$, we can resort to
analytical expressions of the correlation functions for the heavy
baryon currents with one heavy quark  in the finite mass limit
\cite{GrooteF1,GrooteF2} or infinite mass limit
\cite{GrooteI1,GrooteI2} to make possible improvement for the
predicted masses $M_{\Omega_a}$, while the  next-to-leading order
corrections to the perturbative terms  for the baryon currents with
three massless quarks are calculated in
Refs.\cite{massless1,massless2,massless3,massless4}. The analytical
expressions of the perturbative $\alpha_s$ corrections  presented in
Ref.\cite{GrooteF1} are lengthy enough; while the expressions in the
heavy quark limit $m_Q\rightarrow \infty$ are simpler and more easy
to   deal  with, however, the interpolating current differs  from
mine remarkably \cite{GrooteI1,GrooteI2}. The mass of the $s$ quark
plays an important role,  the $\Omega_b$ and $\Sigma_b$  have equal
masses  in the  limit $m_s=0$, from the experimental data
$M_{\Sigma_b}=(5807.8\pm2.7)\, \rm{MeV}$ \cite{PDG}, we can see that
the flavor $SU(3)$ breaking effects (about $350\, \rm{MeV}$) are
rather large,  $m_s=m_u=m_d=0$ maybe a crude approximation. We
prefer another work to study those perturbative effects in detail.

\section{Conclusion}
In this article, we employ the QCD sum rules to calculate the masses
and the pole residues
  of the heavy baryons $\Omega_c^0(css)$ and $\Omega_b^-(bss)$
    including the contributions of the vacuum condensates adding
up to dimension six in the operator product expansion. The numerical
values $M_{\Omega_c^0}=(2.72\pm0.18)\,\rm{GeV}$ (or
$M_{\Omega_c^0}=(2.71\pm0.18)\,\rm{GeV}$) and
$M_{\Omega_b^-}=(6.13\pm0.12)\,\rm{GeV}$ (or
$M_{\Omega_b^-}=(6.18\pm0.13)\,\rm{GeV}$) are   in good agreement
with the experimental data, while the existing theoretical
predictions for the mass $M_{\Omega_b^-} $ is about $0.1 \,
\rm{GeV}$ lower than the experimental value. We can take the pole
residues $\lambda_{\Omega_c}$ and $\lambda_{\Omega_b}$ as  basic
parameters and study the hadronic processes, for example,
$\Omega^*_Q \to \Omega_Q \gamma$.

\section*{Acknowledgements}
This  work is supported by National Natural Science Foundation,
Grant Number 10775051, and Program for New Century Excellent Talents
in University, Grant Number NCET-07-0282.

\end{document}